\documentclass[sigconf,anonymous=false]{acmart}

\usepackage{booktabs}
\usepackage{subcaption}
\newcommand{\eat}[1]{}

\usepackage{listings}
\usepackage{xcolor}


\begin{document}

\title{CodeR$^\mathbf{3}$: A GenAI-Powered Workflow Repair and Revival Ecosystem}

\author{Asif Zaman}
\affiliation{%
  \institution{Department of Computer Science}
    \institution{University of Idaho}
  \city{Moscow, Idaho}
  \country{United States}
}
\email{zama4038@vandals.uidaho.edu}

\author{Kallol Naha}
\affiliation{%
  \institution{Department of Computer Science}
    \institution{University of Idaho}
  \city{Moscow, Idaho}
  \country{United States}
}
\email{naha7197@vandals.uidaho.edu}

\author{Khalid Belhajjame}
\affiliation{%
  \institution{PSL, LAMSADE}
    \institution{Universit\'e Paris-Dauphine}
  \city{Paris}
  \country{France}
}
\email{khalid.belhajjame@dauphine.fr}

\author{Hasan M. Jamil}
\affiliation{%
  \institution{Department of Computer Science}
    \institution{University of Idaho}
  \city{Moscow, Idaho}
  \country{United States}
}
\email{jamil@uidaho.edu}

\renewcommand{\shortauthors}{Zaman, Naha, Belhajjame and Jamil}

\begin{abstract}
Scientific workflows encode valuable domain expertise and computational methodologies. Yet studies consistently show that a significant proportion of published workflows suffer from decay over time. This problem is particularly acute for legacy workflow systems like Taverna, where discontinued services, obsolete dependencies, and system retirement render previously functional workflows unusable. We present a novel legacy workflow migration system, called CodeR$^3$ (stands for Code Repair, Revival and Reuse),  that leverages generative AI to analyze the characteristics of decayed workflows, reproduce them into modern workflow technologies like Snakemake and VisFlow \cite{NahaJ25}. Our system additionally integrates stepwise workflow analysis visualization, automated service substitution, and human-in-the-loop validation. Through several case studies of Taverna workflow revival, we demonstrate the feasibility of this approach while identifying key challenges that require human oversight. Our findings reveal that automation significantly reduces manual effort in workflow parsing and service identification. However, critical tasks such as service substitution and data validation still require domain expertise. Our result will be a crowdsourcing platform that enables the community to collaboratively revive decayed workflows and validate the functionality and correctness of revived workflows. This work contributes a framework for workflow revival that balances automation efficiency with necessary human judgment.
\end{abstract}

\maketitle

\section{Introduction}

Scientific workflows have existed for decades, first emerging in the late 1990s to early 2000s with the aim of modeling and executing in silico experiments and data analyses. Nevertheless, we observe recently that scientific workflows have taken on even greater significance in large-scale, data-intensive and resource-intensive cloud and high performance computing projects. This elevated role is evidenced by their use in experiments involving vast datasets, complex simulations, or large parameter sweeps. To support reproducibility, scalability, and the efficient utilization of distributed infrastructures, workflows are now frequently used to orchestrate pipelines that span cloud, HPC, and edge resources, often automating data movement, provenance capture, and resource provisioning. This expansion is partly evidenced by a surge of interest in scientific workflows over the last five years. New community initiatives, such as the Workflows Community Initiative\footnote{\url{workflows.community}}, have arisen to promote best practices for workflows as first-class digital objects with metadata, versioning, registries, and FAIR compliance \cite{FAIRFlow2025}. For example, WorkflowHub \cite{workflowhub} has emerged as a domain-agnostic registry for describing, sharing, and publishing workflows under FAIR principles, and projects such as EOSC-Life\footnote{\url{www.eosc-life.eu/}}, ELIXIR\footnote{\url{elixir-europe.org/elixir-topics/workflows}} and FONDA\footnote{\url{fonda.hu-berlin.de/}} have invested in infrastructure to support large-scale, reusable workflows . These developments suggest a growing maturity in workflow management, emphasizing interoperability, reproducibility, and reusability.

Despite considerable progress in the field of scientific workflows, the problem of workflow decay remains a persistent hurdle to reproducibility, reuse, and long-term sustainability. The creation of repositories such as myExperiment \cite{goble2010myexperiment}, and more recently WorkflowHub and nf-core\footnote{\url{nf-co.re/}}, has facilitated workflow sharing and publication, but these platforms do not guarantee long-term preservation, and in many cases workflows become unusable even after a short period.
Indeed, workflow decay manifests through multiple mechanisms that render previously functional workflows inoperable. Analysis of the myExperiment repository revealed that nearly 80\% of tested Taverna workflows failed to execute or reproduce the same results. Workflows from earlier years (2007-2009) showing failure rates exceeding 80\% \cite{zhao2012workflows}. The study identified that 50\% of workflow decay was attributed to volatile third-party resources, including unavailable web services, inaccessible databases, and changes to service implementations. Additional decay factors include missing example data, insufficient execution environments, and incomplete workflow documentation. While this analysis focused specifically on Taverna workflows, the methodology and findings are expected to apply broadly to workflow systems that depend on third-party resources.

Beyond the traditional decay factors, we observe a new category of obsolescence related to the retirement of workflow systems themselves. The myExperiment repository contains thousands of workflows implemented in Taverna, a system that was widely adopted by bioinformaticians from the mid-2000s through approximately 2015. As the community has migrated to newer systems like Snakemake \cite{koster2012snakemake} and Nextflow \cite{di2017nextflow}, Taverna workflows have become functionally obsolete. The official retirement of Taverna by Apache in 2020 has further accelerated this obsolescence, leaving valuable scientific methodologies trapped in an unmaintained execution environment.

In this paper, we argue that many decayed workflows represent valuable intellectual investments that warrant systematic revival efforts. They often encapsulate methodological knowledge refined through years of scientific practice and embody established solutions in computational methodology. If properly preserved and modernized, such workflows could continue to benefit the scientific community and, in many cases, provide insights into how experimental and analytical practices have evolved over time. Therefore, rather than treating decay as inevitable, we advocate for principled approaches to workflow revival that restore functionality while safeguarding the underlying scientific methodology.

\subsection{An Illustrative Example}

\begin{figure}
\centering
\includegraphics[width=0.15\textwidth]{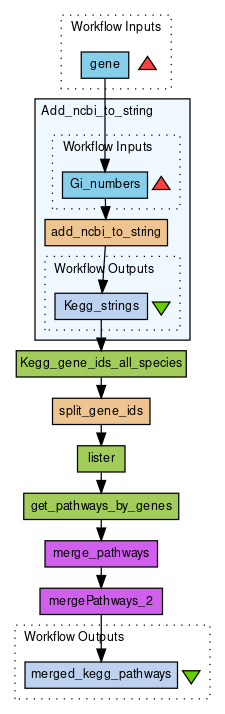}
\caption{Example of a Taverna Workflow for Entrez GeneID to KEGG Pathway Mapping.}
\label{fig:taverna_workflow}
\end{figure}

The simple Taverna workflow shown in Figure \ref{fig:taverna_workflow}, retrieved from the myExperiment workbench, is emblematic of the workflow decay challenges our research aims to address. Mapping gene identifiers to KEGG pathways is a cornerstone of functional genomics and systems biology because it connects individual gene-level signals to higher-order biological processes. For instance, GeneID 7124, which corresponds to TNF (tumor necrosis factor), is mapped to the KEGG pathway hsa05134 (Legionellosis). This association reveals that TNF is not merely a general mediator of inflammation but also a key player in the host immune response to {\em Legionella pneumophila}, the pathogen responsible for Legionnaires’ disease. Situating TNF within this pathway context allows researchers to better understand how molecular events drive cellular defense mechanisms, how pathway dysregulation contributes to disease, and where potential therapeutic interventions might be targeted.

\begin{figure}
\centering
\includegraphics[width=0.49\textwidth]{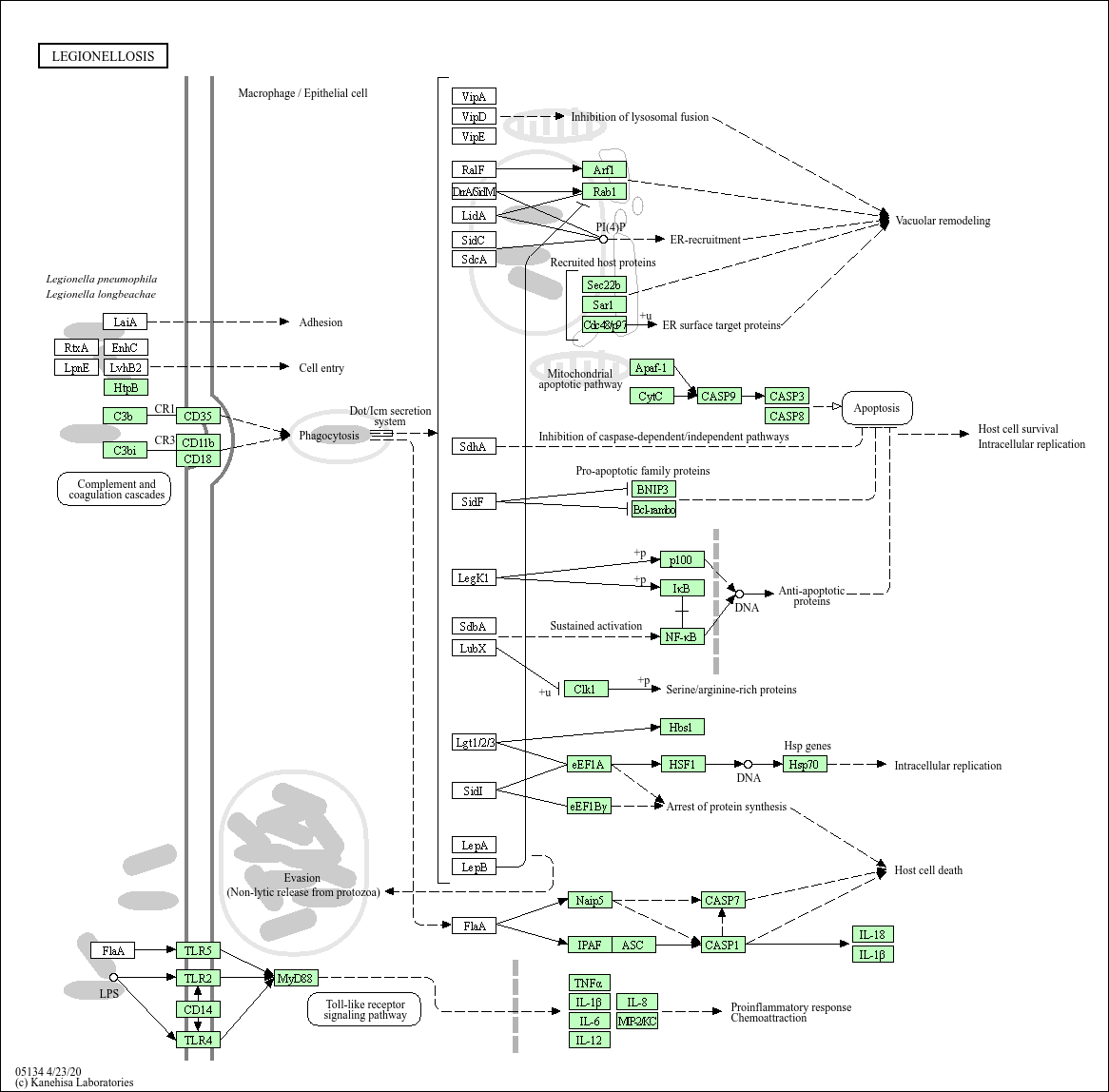}
\caption{KEGG pathway hsa05134 retrieved by CodeR$^3$ generated Snakemake workflow for gene ID 7124.}
\label{fig:kegg}
\end{figure}

Despite its importance, establishing such gene-to-pathway mappings is rarely straightforward. While tools such as the KEGG REST API and KEGG Mapper allow direct database queries, bioinformatics packages like clusterProfiler and KEGGREST (in R) or BioServices (in Python) can automate mapping and enrichment analyses. Yet the challenges become most apparent when examining the lifecycle of older scientific workflows  --  such as the Taverna pipeline in Figure \ref{fig:taverna_workflow}. Originally designed to automate functional annotation and gene-to-pathway analysis, this workflow no longer executes successfully because its service dependencies have decayed: remote web services have been deprecated, APIs have changed or vanished, and underlying data schemas have evolved. These manifestations of “workflow rot” highlight the fragility of computational pipelines that rely on external resources and emphasize the urgent need for modern revival systems like CodeR$^3$. By automatically reinterpreting legacy workflows, updating service calls, and regenerating executable pipelines with contemporary tools (e.g., Snakemake) and databases, CodeR$^3$ enables robust and reproducible workflows, in order to link GeneID 7124 to KEGG pathway hsa05134 as shown in Figure \ref{fig:kegg}.

Examining the spectrum of solutions for revival of legacy workflows such as the ones in Figure \ref{fig:taverna_workflow}, we can identify two main classes. The first includes traditional approaches, where developers work closely with domain experts to manually restore decayed workflows. This is, for example, the approach followed by Belhajjame et al.~\cite{BelhajjameGSR11}, in which a developer attempts to repair a workflow by manually identifying substitute services to replace unavailable ones, often relying on semantic annotations. Such approaches demand substantial effort and cost, in addition to intensive collaboration between domain experts and developers to ensure that the revived workflow faithfully reproduces the experiment or analysis originally targeted by the decayed workflow. Also, when dealing with legacy systems, the process would require expertise in both legacy and modern workflow systems, in addition to a deep understanding of the scientific domain. It is worth noting here that apart from workflow systems, organizations across industries have reported difficulty maintaining legacy computational infrastructure as original developers move on and institutional knowledge dissipates.

The second class, which is the focus of our research, encompasses emerging approaches in which domain experts, supported by large language models (LLMs), directly curate and revive workflows with moderate effort and high quality, without necessarily involving developers in the revival process. The work most closely related to ours is a recent study~\cite{acm-rep2025}, in which the authors outlined a method for workflow repair. While relevant, that contribution is primarily a position paper: the method was not implemented, and the curation process remained largely manual.

\subsection{Our Contributions}

In contrast, our research seeks to test the limits of automation in order to shield domain experts from the hurdles of coding, while ensuring that the revived workflow is both executable and faithful to the original experiment or analysis, modulo necessary modernization. Our solution is embodied in an ecosystem that we call \textit{CodeR$^3$}, with the following main contributions:

\begin{itemize}
\item A systematic methodology for AI-assisted workflow revival that combines automated analysis with human validation to transform legacy workflows into modern executable formats.
\item An evaluation of LLM capabilities and limitations in workflow understanding, component classification, and service substitution identification.
\item A hybrid automation approach that continuously builds a knowledge base from each successfully revived workflow, diagnoses common errors, and constructs more reliable workflows as new workflows are introduced.
\item Taverna case studies, which demonstrated the application of our methodology to revive a decayed Taverna workflow from the myExperiment repository.
\item A framework for community adoption that balances automation efficiency with necessary domain expertise validation.
\end{itemize}

Accordingly, the remainder of this paper is organized as follows. We start by analyzing related work in Section \ref{sec:relatedwork}, and advocating for a hybrid solution that leverages both humans and automation in Section \ref{sec:hybrid}.
We then present our solution in Section \ref{sec:solution} by presenting the underlying methodology and architecture and detailing its component steps.  We report on the results of empirical evaluation that we conducted against A0 real-world decayed workflows in Section \ref{sec:evaluation}, before closing in Section \ref{sec:conclusions}.

\section{Related Work}
\label{sec:relatedwork}

Scientific workflows are fundamental to modern computational research but remain highly vulnerable to decay as software, services, and environments evolve. Existing solutions -- from provenance capture and containerization to automated code repair and LLM-based generation -- address aspects of this challenge but fall short of reviving broken workflows or reconstructing their lost semantics. This section reviews the state of the art and positions CodeR$^3$ as a novel framework that unites generative AI, semantic reconstruction, and community collaboration to enable workflow revival and reuse.

\subsection{Workflow Decay, Preservation, and FAIR Ecosystems}

Workflows encode the logic, dependencies, and provenance of complex analyses, yet their longevity is limited. Studies show that over half of published workflows become non-executable within a few years due to service failures, API changes, data evolution, and dependency conflicts \cite{zhao2012workflows,MaligeayBB24}. Broader evidence highlights similar fragility in computational artifacts -- only 24\% of Jupyter notebooks run without errors, and fewer than 5\% reproduce original outputs \cite{pimentel2019large,trisovic2022large}.

Community efforts have focused on preservation through registries, provenance standards, and environment capture. WorkflowHub treats workflows as shareable, citable assets under FAIR principles \cite{workflowhub,FAIRFlow2025}, and RO-Crate packages execution context and provenance in machine-actionable formats \cite{SoilandReyes2022}. Workflow engines like Snakemake and nf-core promote reproducibility and sustainability via templating and governance \cite{Molder2021,Ewels2020}, while containerization tools such as Docker mitigate environment drift. However, these approaches largely preserve existing workflows rather than repairing broken ones, reconstructing missing semantics, or adapting outdated services.

\subsection{LLMs for Workflow Comprehension and Repair in CodeR$^3$}

LLMs have demonstrated significant capabilities in code generation and automated program repair. Models like Codex achieve 28.8\% accuracy on HumanEval, and others exceed 59\% on synthesis benchmarks such as MBPP \cite{chen2021evaluating,austin2021program}. Automated program repair approaches using LLMs -- evaluated on SWE-bench and related benchmarks -- can localize faults and generate fixes across large repositories \cite{OpenAI2024Introducing,AleithanXMNUW2024,YangCLLZBLT2025}. Yet these methods rely heavily on test oracles, struggle with incomplete specifications, and are typically limited to isolated codebases, failing to capture the orchestration and semantics of full scientific workflows \cite{Zubair2025}.

Recent efforts have applied LLMs to modernization tasks such as legacy translation and documentation synthesis \cite{Busch2025,DiggsOthers2024}, but these remain program-centric and lack end-to-end solutions for workflow recovery. CodeR³ addresses this gap by reconstructing a conceptual, engine-agnostic representation from legacy workflows before generating updated implementations. It handles API drift through LLM-suggested service substitutions, wraps replacements with adapters, and incorporates expert feedback to resolve ambiguities. Crucially, CodeR$^3$ integrates community validation to assess scientific plausibility when no ground truth exists.

By combining semantic reconstruction, service adaptation, and collaborative validation in a single platform, CodeR$^3$ moves beyond preservation or isolated code repair. It operationalizes a human-AI approach to revive decayed workflows, transforming them into reusable, future-ready assets and extending the computational legacy of science.

\section{Need for Human–Machine Collaboration in Code Repair}
\label{sec:hybrid}

LLMs have evolved from offering simple bug fixes to performing repository-scale repairs, localizing faults, generating patches, and iteratively refining solutions across files, dependencies, and APIs \cite{JimenezYWYPPN24,OpenAI2024Introducing}. Yet benchmark results also reveal critical limitations that make human oversight indispensable.

Generalization remains fragile: performance can drop by nearly 50\% when bugs are perturbed, reflecting reliance on superficial patterns rather than semantic understanding \cite{LiJSZ25}. Moreover, test suites are often incomplete or unreliable, so “passes tests” does not guarantee correctness \cite{HuangMPSWM2025}. These issues are magnified in scientific and safety-critical code. For example, translating legacy Fortran kernels into Kokkos C++ demonstrates LLMs’ potential for combined repair and modernization, but numerical stability, concurrency, and performance portability frequently require expert review \cite{GuptaKMSD2025,Godoy2024,Liu2024}.

Hybrid approaches further underscore the value of human involvement. Generating failing-then-passing tests improves patch precision \cite{MundlerMHV24}, and agentic repair systems like RepairAgent integrate planning and environment interaction while relying on developers for ambiguous decisions \cite{Bouzenia2024}. Verification-driven methods combine LLM reasoning with formal proofs, requiring humans to refine specifications. Security studies also highlight vulnerabilities, dependency hallucinations, and unsafe code patterns in AI-generated patches, necessitating layered defenses such as static analysis, curated dependencies, and policy enforcement \cite{Sajadi2025,Zubair2025}.

These findings point to a clear design principle embodied by CodeR$^3$. LLMs are powerful generators of hypotheses -- suggesting fixes, substitutions, and translations -- but robust repair of complex or safety-critical systems demands a tightly coupled human–machine process. Curators supply exemplars, developers guide the search, domain experts encode invariants, and infrastructure delivers strong oracles via testing, tracing, and verification. Systems that integrate these human roles, as CodeR$^3$ does, achieve far greater precision, robustness, and trustworthiness than purely automated repair \cite{MundlerMHV24,HaqueBFV2025,LiJSZ25}.

\section{CodeR$^3$ Architecture}
\label{sec:solution}

\begin{figure*}
\centering
\includegraphics[width=0.85\textwidth]{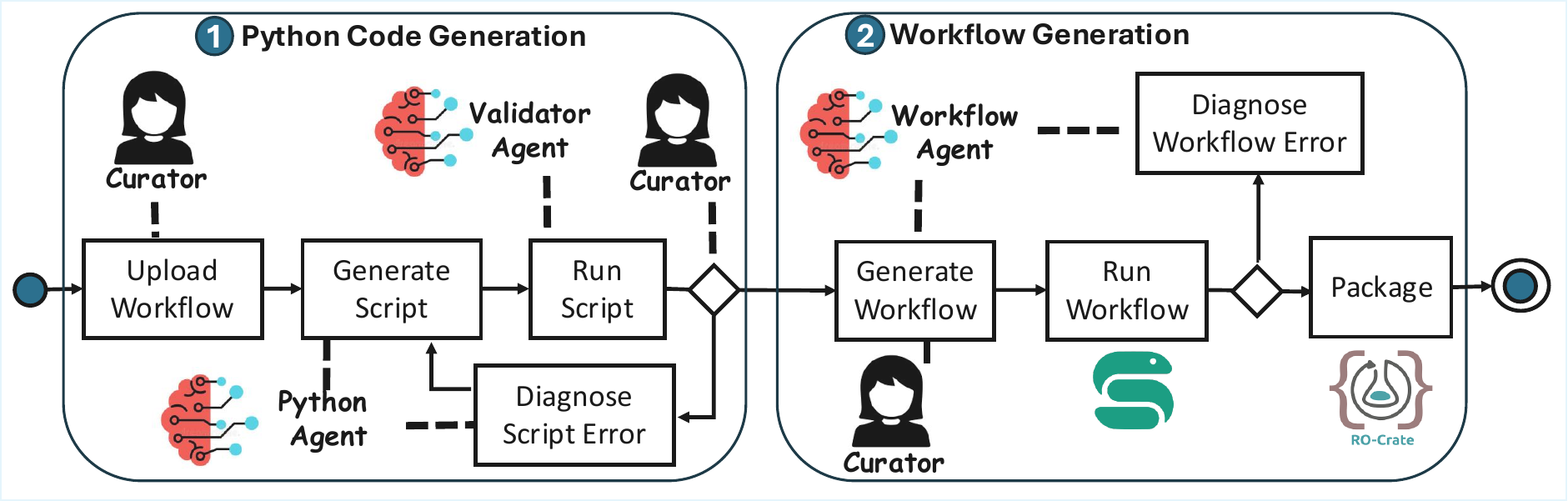}
\caption{Overview of the CodeR$^3$ system and core components}
\label{fig:system_overview}
\end{figure*}

In designing our solution for workflow repair and revival, we sought to satisfy the following guiding requirements. First, the workflow obtained as a result of the revival must be executable. This means that the curator and subsequent users should be able to run the workflow and inspect the results it produces. Second, the revival process must accommodate structural flexibility. Our goal is not merely to reconstruct a workflow with the identical graph structure as the decayed workflow, but to ensure that it implements the same or a similar experiment or analysis as the decayed one- " semantic equivalence preservation". We consider this to be a key requirement because the structure of a workflow is often influenced by the presence of adapters (also called shims) used solely to resolve data type mismatches with little impact on the experiment or analysis at hand. Third, the revival operation must be language-agnostic in that it should not be restricted to a specific workflow language, but instead allow the curator to revive the workflow in the language most suitable for their community of users. Last but not least,  seek feedback from the curator at key decision points to ensure the revised workflow meets expectations. This approach uses the curator as a strategic resource, preserving automation for the majority of the process.

Based on the aforementioned requirements, we have developed a system architecture that serves as the foundation for $CodeR^3$; the operational principles of this system are illustrated in Figure~\ref{fig:system_overview}. The revival operation itself is decomposed into two major sequential steps: the first is \textbf{Python code generation and validation}, and the second is \textbf{workflow generation}. In the sections that follow, we will provide a detailed explanation of each of these core components, and we will systematically justify the specific design choices that were made during their development.

\subsection{CodeR$^3$ Conversational User Interface}
\label{sec:casestudy}

\begin{figure*}
\centering
    \fbox{
\includegraphics[width=.99\textwidth,keepaspectratio]{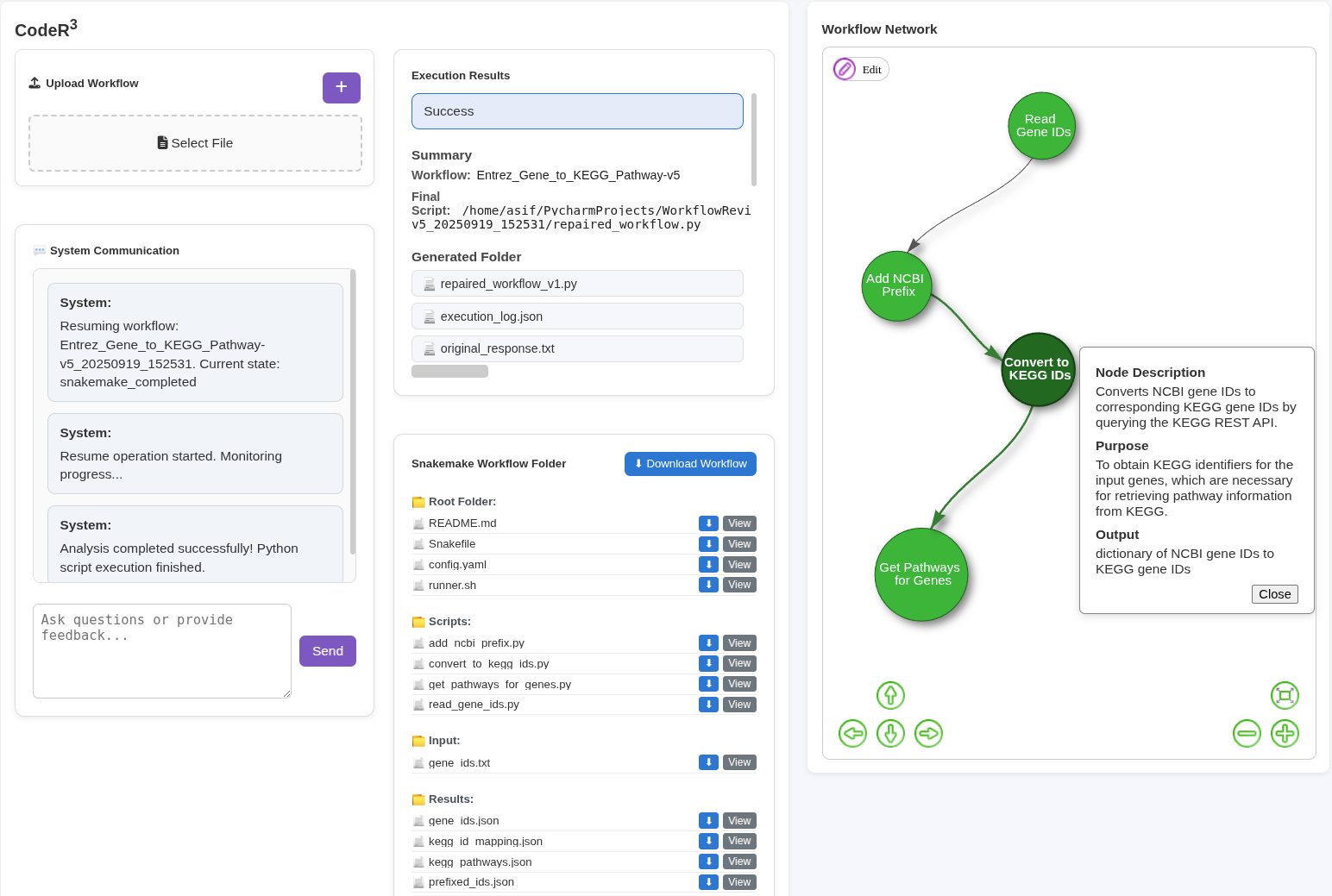}
}
\caption{ The full user interface of the application, displaying Upload, Execution Results, Communication, Network, and Snakemake Workflow sections.}
\Description{The full user interface of the application, displaying Upload, Execution Results, Communication, Network, and Snakemake Workflow sections}
\label{fig:system_screen}
\end{figure*}

Figure \ref{fig:system_screen} shows the GenAI-powered conversational interface of the CodeR$^3$ system, through which users can repair and revive legacy workflows. The interface is organized into three main panels. On the left is the {\em conversation panel}, where users can load a legacy workflow and guide its step-by-step transformation using natural language instructions. The system accepts feedback and directives in plain English and responds conversationally, facilitating an intuitive, interactive repair process.

On the right, the interface presents a {\em graphical representation} of the uploaded workflow, rendered at the highest level of conceptual abstraction that CodeR$^3$ can infer from the underlying code. Each node in this graph represents a distinct functional component, and clicking on a node reveals a conceptual summary of its role within the workflow. Figure \ref{fig:system_screen} illustrates how CodeR$^3$ interprets the Taverna workflow shown previously in Figure \ref{fig:taverna_workflow}, first translating it into an equivalent Python program and then generating a corresponding Snakemake implementation.

Finally, the {\em middle panel} tracks the progress of the conversion process and displays the newly generated workflow components in the user’s chosen workflow management system. This panel also includes an Execute button, which allows users to run the revived workflow directly, examine the output, and validate the accuracy and correctness of the conversion.


\subsection{Python Code Generation and Validation}

The CodeR$^3$ ecosystem is designed to be agnostic to the specific workflow system preferred by the curator and their user community. Consequently, we have chosen to revive decayed workflows into Python scripts rather than targeting any specific workflow language. This design decision offers several advantages: (i) Using Python as an intermediate pivot representation enables support for the full range of processing patterns, dataflows, and control flows that any target workflow language might require; (ii) Debugging a single Python script is more straightforward than debugging within a workflow language environment, as the curator can isolate logic errors from workflow orchestration issues; (iii) Python provides access to extensive libraries for data adaptation and transformation, notably including pandas, which are more comprehensive than those typically available in specialized workflow languages. Listing \ref{lst:service_replacement} depicts the Python code generated by CodeR$^3$ as an intermediate step toward reviving the Taverna workflow in Figure \ref{fig:taverna_workflow}.


Concretely, the CodeR$^3$ ecosystem operates as follows. The curator initiates the process by loading the decayed workflow through a parser capable of handling diverse specification formats. These formats include XML-based workflows (such as Taverna t2flow and SCUFL), programming scripts (including R, JavaScript, and Java), and any associated configuration files. A LLM is subsequently employed to analyze the structural and semantic content of the parsed workflow. The model generates comprehensive Python implementations that preserve the original workflow's scientific methodology while modernizing its execution environment. We emphasize that the generated Python scripts deliberately maintain the modularity of the initial workflows, specifically preserving the distinct component steps. This design is intentional, as it ensures that when the revived workflow is later generated for a target workflow system, the original modular structure can be preserved as much as possible. This is accomplished through a two-stage procedure. First, the system produces a template Python script that systematically maps each component step from the original workflow to a corresponding Python function (module). A main module is simultaneously generated to function as the central orchestrator, managing the execution sequence and data flow between these individual functions. Subsequently, the implementation details for each function body are populated using LLMs. 

\begin{lstlisting}[language=Python,caption={Python implementation showing service substitution process, replacing original SOAP calls with modern REST API endpoints},label={lst:service_replacement}]
....
# URL: http://rest.kegg.jp/conv/genes/ncbi-geneid:7124
for line in response.text.strip().split("\n"):
    if line.strip():
        try:
            ncbi_kegg = line.split("	") #tab seperation
            ncbi_id_returned = ncbi_kegg[0].strip()
            kegg_id = ncbi_kegg[1].strip()

            ncbi_id_short = ncbi_id_returned.split(":")[1] #get only the gene id
            kegg_id_mapping[ncbi_id_short] = kegg_id
......
\end{lstlisting}


The system employs Gemini 2.0 and GPT-5 Mini as the underlying LLM models for this code generation task. These models are guided by a structured prompt template provided in \autoref{lst:minimal_workflow} that instructs them to act as workflow repair specialists. The prompt defines specific requirements: generating a single executable Python file, following a standardized template structure with configuration sections, implementing appropriate branching patterns, and applying service replacement rules for deprecated APIs.

The resulting Python scripts encapsulate the entire workflow functionality. This includes all data processing steps, external service calls, dependency management logic, and output generation mechanisms. This consolidation of functionality into a single, coherent script which can be run and tested for validation. 

The generated Python script undergoes a validation step designed to identify potential issues with workflow execution or output quality. To achieve this, the system interacts with users through a simplified question-answer interface that requires no technical expertise. This human-in-the-loop approach focuses on validating domain knowledge rather than programming or system administration tasks. Specifically, the system presents workflow failures and anomalies in plain language, requesting user feedback regarding expected behaviors, data formats, or scientific interpretations. Users can respond using natural language descriptions without needing knowledge of programming syntax or system configurations. 

\begin{lstlisting}[language=Python,caption={Minimal Workflow Repair Template},label={lst:minimal_workflow}]
You are a workflow repair specialist that converts broken scientific workflows into a single working Python script.
## CORE REQUIREMENTS
### 1. Generate ONE Python file that works immediately
- Complete implementation in single script
- Simple configuration section at top
-.......
### 2. Use This Exact Template Structure:
```python
#!/usr/bin/env python3
"""
Repaired Workflow: {WORKFLOW_NAME}
Original: {ORIGINAL_FORMAT} -> Repaired Python Script  
Domain: {bio/chem/physics/materials/generic}
"""
import ....
CONFIG = {
    'input': 'input/input.FORMAT',
    'output': 'results/output.json', 
    'apis': {'main_service': 'https://api.example.com/',
        'backup_service': 'https://backup-api.com/'}}
def step_1_description():
    """Brief description of what this step does"""
    print(" Step 1: Description")
    # Implementation here
    return result
.....................
```
### 3. Handle Branching with Simple Patterns:
.....................
### 4. Service Replacement Rules:
.....................
\end{lstlisting}

The system also incorporates external resource validation, enabling users to provide alternative service endpoints or data sources when automated discovery fails. This capability proves particularly valuable for workflows that depend on specialized databases or domain-specific services which may not be discoverable through standard registry searches.

\subsection{Workflow Revival, Generation, and Performance}

To automate the end-to-end revival of legacy scientific workflows, beginning with legacy formats such as Taverna XML, CodeR$^3$ parses and reconstructs workflow structures, extracting processors, data links, and service endpoints. This automated analysis accurately recovers the logical flow of the original pipeline and isolates key components needed for execution. One of the most challenging aspects of this process is service substitution: while CodeR$^3$ reliably identifies modern equivalents  --  for instance, replacing deprecated SOAP-based KEGG services with RESTful APIs  --  generating precise endpoint patterns often proves difficult. The model may, for example, propose plausible but incorrect routes such as /convert\_gene instead of the valid /conv endpoint. These implementation-specific details require minimal but critical human-in-the-loop validation to ensure functional correctness.

Once the workflow logic has been reconstructed, CodeR$^3$ generates a syntactically correct and logically coherent Python implementation, abstracting away obsolete platform-specific constructs while aligning with contemporary API usage patterns. This translation step effectively decouples the workflow from legacy infrastructure, enabling downstream portability and extensibility. The validated Python version then serves as the foundation for automated conversion into multiple target workflow systems  --  including Snakemake, Nextflow, and containerized execution environments. During this stage, CodeR$^3$ analyzes data flow patterns, dependency relationships, and resource requirements to produce well-structured rule-based pipelines that faithfully preserve the semantics of the original workflow. A lightweight validation step with the user ensures that the revived workflow behaves as expected before final packaging.

The final phase of the revival pipeline produces complete, self-contained workflow bundles designed for reproducibility and independent execution. These packages include source code, configuration files, sample data, and executable scripts, organized within standardized file hierarchies. Provenance information links the revived workflow to its original version, ensuring transparency and traceability  --  critical elements of reproducible research.

Extensive testing across more than 15 workflow revivals demonstrates the efficiency and robustness of this approach. As summarized in Table \ref{tab:experiment_times_updated}, the complete revival process typically takes 5–10 minutes per workflow, with automation covering 80–90\% of the pipeline. The limited human intervention required is concentrated primarily in domain-specific validation, such as correcting API endpoint patterns or verifying scientific plausibility. Together, these results demonstrate that CodeR$^3$ offers an effective and scalable solution for transforming legacy workflows into modern, reusable computational assets, significantly reducing technical barriers while preserving scientific integrity.

\begin{lstlisting}[language=Python,caption={Generated Snakemake workflow specification with organized rule structure and clear dependency relationships},label={lst:snakemake_workflow}]
configfile: "config.yaml"
os.makedirs("results", ......

rule all:input: config["output"]
rule read_gene_ids:...
rule add_ncbi_prefix:...

rule convert_to_kegg_ids:
    input: "results/prefixed_ids.json"
    output: "results/kegg_id_mapping.json"
    log: "logs/convert_to_kegg_ids.log"
    params: kegg_api=config["kegg_api"]
    script: "scripts/convert_to_kegg_ids.py"
    
rule get_pathways_for_genes:...
\end{lstlisting}









\begin{table}
\centering
\caption{Single File Experiment Execution Times}
\label{tab:experiment_times_updated}
\footnotesize
\begin{tabular}{|p{5cm}|c|c|}
\hline
\textbf{Experiment} & \textbf{Python} & \textbf{Snakemake} \\
\hline \hline
Fetch USE from mets & 52.4s & 20.7s \\
\hline
Entrez Gene to KEGG Pathway-v5 & 40.8s & 28.9s \\
\hline
Generate CombiUgi library-v1 & 16.0s & 16.2s \\
\hline
Download CAS numbers and save as SD file-v1 & 45.7s & 12.57s \\
\hline
Fetch PDB flatfile from RCSB server-v1 & 2.7s & 12.57s \\
\hline
Example of interoperability validation on real time with PDL services-v2 & 2.0s & 24.05s \\
\hline
Use of rest services described with PDL-v1 & 62.9s & 36.88s \\
\hline
Analysing workflows-v3 & 55.33 & 69.77 \\
\hline
DNA sequence analysis pilot -- Blat -v2 & 4.3s & --- \\
\hline
XPath From VOTable-v1 & 378.2s & 42.14s \\
\hline
BioAID DiseaseDiscovery RatHumanMouseUniprotFilter & 77.08s & 47.61s \\
\hline
BioAID ProteinDiscovery filterOnHumanUniprot perDoc html & 5.93s & 40.53s \\
\hline
BiomartAndEMBOSSAnalysis-v4 & --- & --- \\
\hline
EBI InterProScan-v3 & --- & --- \\
\hline
Nucleotide InterProScan-v4 & 723.24 & 61.61 \\
\hline
Workflow for Protein Sequence Analysis & 840.15 & 126.33 \\
\hline
\end{tabular}
\end{table}

\section{Experimental Evaluation and Results}
\label{sec:evaluation}

We evaluated CodeR$^3$’s automated workflow revival capabilities on ten legacy workflows drawn from bioinformatics and chemistry, with over fifty iterative refinements of key examples to assess performance and robustness. These workflows originated from diverse legacy systems  --  including Taverna .t2flow files, XML specifications, and JavaScript-based definitions  --  and ranged from simple three-step pipelines to complex, multi-stage analyses involving iterative processes, branching logic, and external service dependencies.

Overall, the system demonstrated strong performance across most revival stages. Automated Python script generation achieved over 90\% success in producing syntactically correct code, reflecting the extensive programming knowledge captured by LLMs. Conversion of revived workflows into Snakemake formats achieved approximately 80\% success, yielding executable pipelines with minimal human intervention. Structural parsing of original workflow specifications was consistently reliable, accurately identifying components, data dependencies, and service endpoints across formats.

Workflow complexity strongly influenced success rates. Simple, linear workflows with well-defined data flow and minimal branching were revived almost flawlessly. More complex workflows  --  particularly those having more than 5 steps, nested loops, conditional logic, or intricate interdependencies  --  remained challenging. These difficulties mirror broader limitations of current LLMs in reasoning over extended logical sequences and maintaining global workflow state. In some cases, conditional constructs absent in the original Taverna workflows emerged during automated code generation, underscoring how language models may introduce unintended complexity during translation.

While LLMs reliably identified conceptually equivalent services to replace deprecated endpoints, they often generated plausible but incorrect URL patterns. These inaccuracies required manual correction and validation against updated documentation, particularly as evolving APIs frequently introduced new parameter structures and authentication protocols.

Validation strategies necessarily emphasized scientific plausibility over strict reproducibility, given that definitive ground truth was unavailable for most legacy workflows. Human experts evaluated whether outputs were scientifically reasonable and consistent with expected patterns. Although the system could reliably confirm that revived workflows executed without errors, distinguishing valid scientific outputs from flawed results required domain knowledge. Human input was also crucial in verifying endpoint functionality and preparing suitable input data when original test datasets were missing.

In summary, CodeR$^3$ achieves a high degree of automation in reviving legacy workflows, with strong performance in code generation, structural parsing, and functional transformation. However, service integration, complex control flows, and scientific validation still require human expertise, highlighting the ongoing need for tight human–AI collaboration. These results underscore both the promise and the current boundaries of automated workflow revival, setting the stage for future improvements in reasoning, validation, and adaptive service discovery.

\section{Discussion and Concluding Remarks}
\label{sec:conclusions}

This work presented CodeR$^3$, a generative AI–driven framework for reviving obsolete scientific workflows  --  a growing challenge as APIs evolve, services disappear, and computational environments change. By combining LLM-based automation with human expertise, CodeR³ transforms legacy workflows into modern, executable forms while preserving their scientific intent.

A key difficulty in workflow revival is service discovery and API evolution. While LLMs can often identify functional replacements for deprecated services, they frequently generate plausible but incorrect endpoints, highlighting the need for human oversight. Differences in documentation quality and service maturity across domains further complicate reliable automation.

Validation poses another major challenge. A workflow that runs successfully may still yield scientifically invalid results. Detecting such “silent failures” requires domain knowledge, as evolving databases, updated algorithms, and corrected datasets mean that strict reproducibility is not always possible. Instead, achieving functional equivalence  --  reproducing the original analytical purpose  --  is often a more meaningful goal.

Despite these challenges, CodeR$^3$ reduces the technical barriers to workflow revival, enabling collaborative efforts where AI handles automation and experts focus on validation and interpretation. Our evaluations show that CodeR³ can convert workflows like those built in Taverna into Python or Snakemake implementations with 85–90\% automation coverage, often within minutes. Complex workflows involving loops, conditionals, or intricate dependencies remain difficult, indicating the need for more advanced semantic understanding.

Future work will focus on improving service discovery, automating validation to detect silent failures, and integrating agentic features that retrieve domain-specific resources autonomously. Ultimately, CodeR$^3$ demonstrates that large-scale, automated workflow revival is achievable. By turning static, decayed workflows into living, reusable assets, it strengthens reproducibility, preserves scientific knowledge, and extends the utility of computational research infrastructure well into the future.

\begin{acks}
This research was supported in part by the National Institutes of Health IDeA grant P20GM103408 and the National Science Foundation CSSI grant OAC 2410668.
\end{acks}

\bibliographystyle{plainnat}
\bibliography{references}

@inproceedings{NahaJ25,
  author       = {Kallol Naha and
                  Hasan M. Jamil},
  title        = {Abstract Visual Scientific Workflow Design using VisFlow 2.0},
  booktitle    = {{ACM SSDBM} 2025, Columbus, OH, USA, June 23-25, 2025},
  pages        = {15:1--15:6},
  publisher    = {{ACM}},
  year         = {2025},
}

@inproceedings{BelhajjameGSR11,
  author       = {Khalid Belhajjame and
                  Carole A. Goble},
  title        = {Fostering Scientific Workflow Preservation through Discovery of Substitute
                  Services},
  booktitle    = {{IEEE}  E-Science,
                  Stockholm, Sweden, December 5-8, 2011},
  pages        = {97--104},
  year         = {2011},
}

@article{acm-rep2025,
place = {Country unknown/Code not available}, title = {Resurrecting the Dead: {GenAI}-Powered Workflow Revival}, url = {https://par.nsf.gov/biblio/10631959},  journal = {ACM Rep}, publisher = {ACM}, author = {Belhajjame, K and Jamil, H}, pages = {1-4}, year = 2025}

@article{workflowhub,
  title={WorkflowHub: a registry for computational workflows},
  author={Gustafsson, Ove Johan Ragnar and Wilkinson, Sean R and Bacall, Finn and others},
  journal={Scientific Data},
  volume={12},
  number={1},
  pages={837},
  year={2025},
  publisher={Nature Publishing Group UK London}
}

@article{FAIRFlow2025,
  title={Applying the FAIR principles to computational workflows},
  author={Wilkinson, Sean R and Aloqalaa, Meznah and  others},
  journal={Scientific Data},
  volume={12},
  number={1},
  pages={328},
  year={2025},
  publisher={Nature}
}

@article{Molder2021,
  title={Sustainable data analysis with Snakemake},
  author={M{\"o}lder, Felix and Jablonski, Kim Philipp and  others},
  journal={F1000Research},
  volume={10},
  pages={33},
  year={2021}
}

@article{di2017nextflow,
  title={Nextflow enables reproducible computational workflows},
  author={Di Tommaso, Paolo and Chatzou, Maria and Floden, Evan W and Barja, Pablo Prieto and Palumbo, Emilio and Notredame, Cedric},
  journal={Nature biotechnology},
  volume={35},
  number={4},
  pages={316--319},
  year={2017},
  publisher={Nature Publishing Group US New York}
}

@inproceedings{MaligeayBB24,
  author       = {Nathan Maligeay and
                  No{\'{e}}mie Bossut and
                  Khalid Belhajjame},
  editor       = {Shadi Ibrahim and
                  Suren Byna and
                  Tristan Allard and
                  Jay F. Lofstead and
                  Amelie Chi Zhou and
                  Tassadit Bouadi and
                  Jalil Boukhobza and
                  Diana Moise and
                  C{\'{e}}dric Tedeschi and
                  Jean Luca Bez},
  title        = {Why Do Scientific Workflows Still Break?},
  booktitle    = {Proceedings of the 36th International Conference on Scientific and
                  Statistical Database Management, {SSDBM} 2024, Rennes, France, July
                  10-12, 2024},
  pages        = {15:1--15:4},
  publisher    = {{ACM}},
  year         = {2024},
  url          = {https://doi.org/10.1145/3676288.3676300},
  doi          = {10.1145/3676288.3676300},
  bibsource    = {dblp computer science bibliography, https://dblp.org}
}

@article{chen2021evaluating,
    title = {Evaluating large language models trained on code},
    author = {Chen, Mark and Tworek, Jerry and  others},
    journal = {Corr},
    volume = {arXiv:2107.03374},
    year = {2021}
}

@article{koster2012snakemake,
    title = {Snakemake—a scalable bioinformatics workflow engine},
    author = {K{\"o}ster, Johannes and Rahmann, Sven},
    journal = {Bioinformatics},
    volume = {28},
    number = {19},
    pages = {2520--2522},
    year = {2012},
    publisher = {Oxford University Press}
}

@article{austin2021program,
    title = {Program synthesis with large language models},
    author = {Austin, Jacob and Odena, Augustus and others},
    journal = {arXiv preprint arXiv:2108.07732},
    year = {2021}
}

@article{goble2010myexperiment,
    title = {myExperiment: a repository and social network for the sharing of bioinformatics workflows},
    author = {Goble, Carole A and Bhagat, Jiten and Aleksejevs, Sergejs and Cruickshank, Don and Michaelides, Danius and Newman, David and Borkum, Mark and Bechhofer, Sean and Roos, Marco and Li, Peter and others},
    journal = {Nucleic acids research},
    volume = {38},
    number = {suppl\_2},
    pages = {W677--W682},
    year = {2010},
    publisher = {Oxford University Press}
}

@inproceedings{pimentel2019large,
    title = {A large-scale study about quality and reproducibility of jupyter notebooks},
    author = {Pimentel, Jo{\~a}o Felipe and Murta, Leonardo and Braganholo, Vanessa and Freire, Juliana},
    booktitle = {IEEE/ACM (MSR)},
    pages = {507--517},
    year = {2019},
    organization = {IEEE}
}

@article{trisovic2022large,
  title={A large-scale study on research code quality and execution},
  author={Trisovic, Ana and Lau, Matthew K and Pasquier, Thomas and Crosas, Merc{\`e}},
  journal={Scientific Data},
  volume={9},
  number={1},
  pages={60},
  year={2022},
  publisher={Nature Publishing Group UK London}
}

@inproceedings{zhao2012workflows,
  title={Why workflows break -- Understanding and combating decay in Taverna workflows},
  author={Zhao, Jun and Gomez-Perez, Jose Manuel and Belhajjame, Khalid and Klyne, Graham and Garcia-Cuesta, Esteban and Garrido, Aleix and Hettne, Kristina and Roos, Marco and De Roure, David and Goble, Carole},
  booktitle={IEEE e-Science},
  pages={1--9},
  year={2012},
  organization={IEEE}
}

@article{SoilandReyes2022,
  title = {Packaging research artefacts with RO-Crate},
  volume = {5},
  number = {2},
  journal = {Data Science},
  publisher = {SAGE Publications},
  author = {Soiland-Reyes,  Stian and Sefton,  Peter and Others},
  year = {2022},
  month = jan,
  pages = {97–138}
}

@article{Ewels2020,
  title = {The nf-core framework for community-curated bioinformatics pipelines},
  volume = {38},
  number = {3},
  journal = {Nature Biotechnology},
  publisher = {Springer},
  author = {Ewels,  Philip A. and Peltzer,  Alexander and Fillinger,  Sven and Patel,  Harshil and Alneberg,  Johannes and Wilm,  Andreas and Garcia,  Maxime Ulysse and Di Tommaso,  Paolo and Nahnsen,  Sven},
  year = {2020},
  month = feb,
  pages = {276–278}
}

@misc{OpenAI2024Introducing,
  author = {OpenAI},
  title = {Introducing SWE-bench Verified},
  year = {2024},
  howpublished = {OpenAI. https://openai.com/index/introducing-swe-bench-verified/}
}

@article{AleithanXMNUW2024,
  author       = {Reem Aleithan and
                  Haoran Xue and
                  Mohammad Mahdi Mohajer and
                  Elijah Nnorom and
                  Gias Uddin and
                  Song Wang},
  title        = {SWE-Bench+: Enhanced Coding Benchmark for {LLM}s},
  journal      = {CoRR},
  volume       = {abs/2410.06992},
  year         = {2024},
}

@article{Zubair2025,
  title = {The use of large language models for program repair},
  volume = {93},
  journal = {Computer Standards \& Interfaces},
  publisher = {Elsevier BV},
  author = {Zubair,  Fida and Al-Hitmi,  Maryam and Catal,  Cagatay},
  year = {2025},
  month = apr,
  pages = {103951}
}

@article{YangCLLZBLT2025,
  author       = {Boyang Yang and
                  Zijian Cai and
                  Fengling Liu and
                  Bach Le and
                  Lingming Zhang and
                  Tegawend{\'{e}} F. Bissyand{\'{e}} and
                  Yang Liu and
                  Haoye Tian},
  title        = {A Survey of {LLM}-based Automated Program Repair: Taxonomies, Design
                  Paradigms, and Applications},
  journal      = {CoRR},
  volume       = {abs/2506.23749},
  year         = {2025},
}

@article{DiggsOthers2024,
  author       = {Colin Diggs and
                  Michael Doyle and
                  Others},
  title        = {Leveraging {LLM}s for Legacy Code Modernization: Challenges and Opportunities
                  for {LLM}-Generated Documentation},
  journal      = {CoRR},
  volume       = {abs/2411.14971},
  year         = {2024},
}

@article{Busch2025,
  title = {{LLM}-based code generation and system migration in language-driven engineering},
  volume = {27},
  ISSN = {1433-2787},
  number = {1},
  journal = {Intl. J. on Soft. Tools for Tech. Trans.},
  publisher = {Springer},
  author = {Busch,  Daniel and Bainczyk,  Alexander and Smyth,  Steven and Steffen,  Bernhard},
  year = {2025},
  month = feb,
  pages = {137–147}
}

@inproceedings{JimenezYWYPPN24,
  author       = {Carlos E. Jimenez and
                  John Yang and
                  Alexander Wettig and
                  Shunyu Yao and
                  Kexin Pei and
                  Ofir Press and
                  Karthik R. Narasimhan},
  title        = {SWE-bench: Can Language Models Resolve Real-world Github Issues?},
  booktitle    = {
                  {ICLR} 2024, Vienna, Austria, May 7-11, 2024},
  publisher    = {OpenReview.net},
  year         = {2024},
}

@inproceedings{LiJSZ25,
  author       = {Fengjie Li and
                  Jiajun Jiang and
                  Jiajun Sun and
                  Hongyu Zhang},
  title        = {Evaluating the Generalizability of {LLM}s in Automated Program Repair},
  booktitle    = {{IEEE/ACM} {ICSE} 2025 - NIER, Ottawa, ON, Canada,
                  April 27 - May 3, 2025},
  pages        = {91--95},
  publisher    = {{IEEE}},
  year         = {2025},
}

@article{HuangMPSWM2025,
  author       = {Li Huang and
                  Ilgiz Mustafin and
                  Marco Piccioni and
                  Alessandro Schena and
                  Reto Weber and
                  Bertrand Meyer},
  title        = {Do {AI} models help produce verified bug fixes?},
  journal      = {CoRR},
  volume       = {abs/2507.15822},
  year         = {2025},
}

@misc{GuptaKMSD2025,
      title={From Legacy Fortran to Portable Kokkos: An Autonomous Agentic AI Workflow}, 
      author={Sparsh Gupta and Kamalavasan Kamalakkannan and Maxim Moraru and Galen Shipman and Patrick Diehl},
      year={2025},
      eprint={2509.12443},
      archivePrefix={arXiv},
      primaryClass={cs.SE},
      url={https://arxiv.org/abs/2509.12443}, 
}

@article{Godoy2024,
  title = {Large language model evaluation for high‐performance computing software development},
  volume = {36},
  number = {26},
  journal = {Conc. \& Comp.: Prac. and Exp.},
  publisher = {Wiley},
  author = {Godoy,  William F. and Valero‐Lara,  Pedro and Teranishi,  Keita and Balaprakash,  Prasanna and Vetter,  Jeffrey S.},
  year = {2024},
  month = sep 
}

@article{Liu2024,
  title = {An Empirical Study of the Code Generation of Safety-Critical Software Using {LLM}s},
  volume = {14},
  number = {3},
  journal = {Applied Sciences},
  publisher = {MDPI AG},
  author = {Liu,  Mingxing and Wang,  Junfeng and Lin,  Tao and Ma,  Quan and Fang,  Zhiyang and Wu,  Yanqun},
  year = {2024},
  month = jan,
  pages = {1046}
}

@inproceedings{MundlerMHV24,
  author       = {Niels M{\"{u}}ndler and
                  Mark Niklas M{\"{u}}ller and
                  Jingxuan He and
                  Martin T. Vechev},
  title        = {SWT-Bench: Testing and Validating Real-World Bug-Fixes with Code Agents},
  booktitle    = {NeurIPS 2024, Vancouver,
                  BC, Canada, December 10 - 15, 2024},
  year         = {2024},
}

@misc{Bouzenia2024,
      title={RepairAgent: An Autonomous, {LLM}-Based Agent for Program Repair}, 
      author={Islem Bouzenia and Premkumar Devanbu and Michael Pradel},
      year={2024},
      eprint={2403.17134},
      archivePrefix={arXiv},
      primaryClass={cs.SE},
}

@article{Sajadi2025,
  title = {Do {LLM}s consider security? an empirical study on responses to programming questions},
  volume = {30},
  number = {4},
  journal = {Empirical Software Engineering},
  publisher = {Springer Science and Business Media LLC},
  author = {Sajadi,  Amirali and Le,  Binh and Nguyen,  Anh and Damevski,  Kostadin and Chatterjee,  Preetha},
  year = {2025},
  month = apr 
}

@article{HaqueBFV2025,
  author       = {Mirazul Haque and
                  Petr Babkin and
                  Farima Farmahinifarahani and
                  Manuela Veloso},
  title        = {Towards Effectively Leveraging Execution Traces for Program Repair
                  with Code {LLM}s},
  journal      = {CoRR},
  volume       = {abs/2505.04441},
  year         = {2025},
}

\end{document}